\documentclass[twocolumn, aps, prl]{revtex4-1}
\usepackage{hyperref}
\usepackage{amsfonts}
\usepackage{physics}
\usepackage{lmodern}
\usepackage{xcolor}
\usepackage[utf8]{inputenc}
\usepackage{url}
\usepackage{chngcntr}
\counterwithout{equation}{section}
\hypersetup{
colorlinks = true,
citecolor = magenta,
linkcolor = blue,
urlcolor = violet,
pdftitle={Upamanyu Moitra: Duality-Invariant Higher-Derivative Corrections to Charged Stringy Black Holes},
pdfauthor={Upamanyu Moitra}
}

\usepackage[T1]{fontenc}
\usepackage{graphicx}
\definecolor{refkey}{gray}{0.45}
\definecolor{labelkey}{RGB}{255,0,0}

\newcommand{\ldc}{\xi}
\newcommand{\eig}{\rho}
\newcommand{\mta}{\gamma}
\newcommand{\mtb}{\chi}
\newcommand{\funa}{\beta}
\newcommand{\fe}{\mathfrak{e}}
\newcommand{\fq}{\mathfrak{q}}
\newcommand{\fv}{\mathfrak{v}}

\newcommand{\bee}{\begin{equation}}
\newcommand{\ee}{\end{equation}}
\newcommand{\beq}{\begin{equation*}}
\newcommand{\eeq}{\end{equation*}}
\newcommand{\baa}{\begin{equation}\begin{aligned}}
\newcommand{\ea}{\end{aligned}\end{equation}}

\begin{document}
\title{Duality-Invariant Higher-Derivative Corrections to Charged Stringy Black Holes}
\author{Upamanyu Moitra}
\email{u.moitra@uva.nl}
\affiliation{Institute for Theoretical Physics, Institute of Physics,  Universiteit van Amsterdam, Science Park 904,  1098 XH Amsterdam, The Netherlands}

\begin{abstract}
We study duality-invariant higher-derivative corrections to the charged black hole geometry in two-dimensional heterotic string theory. We illustrate how the conventional perturbative approach to determine the corrected geometry breaks down.  Using a non-perturbative (in $\alpha'$) parametrization of the solution,  we find the corrected charge-to-mass ratio for extremal black holes.  We remark on the results in relation to the weak gravity conjecture.  We also consider the entropy of the extremal black hole within the attractor mechanism and find that the two-derivative entropy is not renormalized to any order. We make comments on interpretations of the results and their extension to near-extremal black holes.
\end{abstract}

\maketitle

\section{Introduction}\label{sec-intro}

Black holes (BHs) are essential objects in our efforts to understand various features of quantum gravity (QG).  String theory,  a leading candidate for a unified theory of all known interactions, has elucidated many aspects of QG.  While general questions involving BHs in string theory can be intractably difficult to answer,  a consideration of lower spacetime dimensions often simplifies computations, while offering insights about general aspects of gravity.   Even in low dimensions, however, it is extremely important to take into account the straitjackets that make string theory compelling as a theory of QG.  Among such constraints,  target space duality or $T$-duality \cite{Giveon:1994fu},  describing the equivalence of physics on small and large circles, is particularly important.  It is known that at tree-level,  strings compactified on a $d$-dimensional torus have $\mathrm{O}(d,d; \mathbb{R})$ as an exact duality symmetry group \cite{Meissner:1991zj, *Sen:1991zi, *Gasperini:1991ak, *Maharana:1992my}.

In the recent years,  there has been significant progress in understanding duality-symmetric theories with double field theory-based  methods \cite{Siegel:1993th,  *Hull:2009mi}.  Inspired by early work \cite{Meissner:1996sa},  there now exists a complete classification of perturbative higher-derivative (HD) corrections for cosmological and $1+1$-dimensional static backgrounds \cite{Hohm:2019jgu, *Codina:2023fhy} in the massless sector of closed string theory.  In recent work \cite{Moitra:2025fhx},  we extended this classification program to include heterotic strings and the associated gauge fields, truncated to the Cartan subgroup of the gauge group.  For a single Maxwell field, the duality group is $\mathrm{O}(1,2; \mathbb{R})$. The extension of the formalism in \cite{Moitra:2025fhx} allows for a detailed study of two-dimensional charged BHs in the heterotic theory,  first discovered in \cite{McGuigan:1991qp},  and various interesting associated questions.  We also showed in \cite{Moitra:2025fhx} how a non-perturbative parametrization \cite{Gasperini:2023tus,  Codina:2023nwz} of the solution in the presence of arbitrary HD terms could be extended to the heterotic case.

The goal of the present work is to study some physically interesting aspects of charged BHs within this framework.  Charged BHs with degenerate horizons are known as \emph{extremal}.  Extremal BHs have played a crucial role in  illuminating various aspects of QG, from microscopic state-counting \cite{Strominger:1996sh} to the ongoing efforts of mapping the landscape of consistent QG theories \cite{Vafa:2005ui},  known as the \emph{Swampland} program.  One statement of the \emph{weak gravity conjecture} (WGC) \cite{Arkani-Hamed:2006emk},  a central part of this program,  is that extremal BHs should be able to decay by emitting particles of sufficiently high charge-to-mass ratio.  It was also suggested \cite{Kats:2006xp} that BHs themselves could meet the criterion of such particles: consistent HD corrections in the low-energy effective field theory (EFT) were argued to \emph{increase} the extremal charge-to-mass ratio.
This has been explicitly checked in several scenarios \cite{Cano:2019oma,  *Cano:2019ycn,  *Cano:2021nzo}.

One motivation of this work is addressing such questions in the two-dimensional framework. There are many known subtleties related to the Swampland program  in $d=2$ \cite{McNamara:2020uza}.   For instance,  the \emph{no-global-symmetries} conjecture is believed to be generally valid only for $d\geq4$,  when gravity is dynamical.  In $d=2$,  both gravity and electromagnetism are non-dynamical and the electrostatic potential for a point charge grows linearly with distance.  A general statement of WGC cannot be easily extended to $d=2,3$; see \cite{Moitra:2023yyc}.  Therefore, it is not \textit{a priori} clear what a Swampland-like statement would resemble and whether one can make any such universal statement at all. 
 On the other hand,  the BH we consider is a genuine string background arising from a world-sheet description.  It has been argued \cite{Banks:1988yz} that continuous global symmetries on the world-sheet always translate to gauge symmetries in spacetime.  Given this, one does expect an absence of global symmetries in the background and consequently,  something akin to WGC. We hope that this work is the first step towards addressing such subtle questions.
 
 In what follows,  we explore various physical consequences of duality-invariant HD corrections.  First, we calculate the corrected extremal charge-to-mass ratio.  While part of the motivation was outlined in the previous paragraph,  this calculation, not previously attempted in the literature, is of broader interest in BH physics.  We first perform the computation in perturbation theory with only the four-derivative terms and encounter a surprising difficulty.  To circumvent this,  we use a non-perturbative formalism \cite{Gasperini:2023tus,  Codina:2023nwz, Moitra:2025fhx}. Pleasingly,  we are able to write down an explicit formula for the ratio,  with a remarkable feature, which we comment upon.
 
 We next turn our attention to the calculation of Wald entropy \cite{Wald:1993nt} of the extremal BH with the HD terms within the attractor mechanism \cite{Ferrara:1995ih}. To this end, we apply the entropy function formalism \cite{Sen:2005wa, *Sen:2007qy} proposed by Sen.  We find yet another striking result concerning the non-renormalization of the two-derivative entropy.  We conclude the Letter with some comments on interpretations of the foregoing results,  BHs slightly away from extremality and possible future directions. We have relegated some formulas to the End Matter on pp.\pageref{sec-supp}.

\section{Corrected Extremal Charge-to-Mass Ratio}\label{sec-corrqm}
The two-derivative target space action describing the massless sector of heterotic strings involving the metric $g_{\mu\nu}$,  string dilaton $\phi$ and Maxwell field $A_\mu$ is
\baa
I_{(2)} = \int \dd[2]{x} \sqrt{-g} e^{-2\phi}\bqty{ \ldc^2 + R + 4(\nabla \phi)^2 -\frac{F_{\mu \nu} F^{\mu \nu} }{4}  }, \label{2derac}
\ea
where $\ldc^2\equiv8/\alpha'$,  $\sqrt{\alpha'}$ being the string length.  We have absorbed the Newton constant in $\phi$.
The full action is augmented with HD terms,  involving combinations of $R$, $F_{\mu\nu}$ and $\nabla_\mu\phi$ and their derivatives.  While the general HD action is complicated in form,  enormous simplifications occur under the assumption of duality in a time-independent background described by
\baa
\dd{s}^2&=-m(x)^2\dd{t}^2+n(x)^2\dd{x}^2, \,  \\
\phi &= \phi(x), \quad A_\mu = V(x) \delta^t_\mu. \label{ansata}
\ea
 Defining the duality-invariant dilaton 
 \baa 
 \Phi(x)=2\phi(x)-\log|m(x)|,  \label{dinvdil}
 \ea 
 under appropriate field redefinitions \cite{Hohm:2019jgu,  *Codina:2023fhy, Moitra:2025fhx},  the full action, dimensionally reduced,  can be shown to assume a remarkably simple $\mathrm{O} (1, 2; \mathbb{R})$-invariant form,
\baa
I_{\rm full} = \int \dd{x} n(x) e^{-\Phi} \pqty{\ldc^2 + \frac{\Phi'(x)^2}{n(x)^2} + F\qty(\eig)  }, \label{actfull}
\ea
 where $F(\eig)$ is a function of the \emph{single} variable $\eig$,  a duality-invariant combination of the curvature and gauge field-strength,
\baa
F(\eig) = -\frac{1}{2} \eig^2 + \sum_{n = 2}^\infty c_{2n} \eig^{2n}, \quad \eig^2 \equiv \frac{2m'(x)^2 -V'(x)^2}{n(x)^2 m(x)^2}. \label{feigs}
\ea
We can generalize the discussion to include $r$ abelian gauge fields,  leading to the symmetry group $\mathrm{O} (1, 1+r; \mathbb{R})$. Remarkably,  duality is so constricting that there is \emph{only one} Wilson coefficient at each derivative order \cite{Moitra:2025fhx}.  
For definiteness,  we consider a single Maxwell field. 

The two-derivative action admits the well-known charged BH solution \cite{McGuigan:1991qp} 
\baa
\overline{m} (x)^2 &= 1 - 2\mu e^{-\ldc x} + Q^2 e^{-2\ldc x} = \overline{n}(x)^{-2}, \\
\overline{\phi} (x)  &= - \frac{1}{2} \ldc x, \qquad \overline{V} (x) = - \sqrt{2} Q e^{-\ldc x}.\label{2dersol}
\ea
The physical mass and charge are $2\ldc\mu$ and $\sqrt{2}\ldc Q$ respectively but we will keep using the parameters $\mu$ and $Q$. 
The BH is non-extremal for $\mu > |Q|$ and extremal for $\mu = |Q|$.  Henceforth, we consider a positive $Q$. The BH has horizons at $x_\pm=\ldc^{-1}\log(\mu\pm\sqrt{\mu^2-Q^2})$.  

HD terms are known to modify such a linear mass-charge relationship,  characteristic of asymptotically flat extremal BHs in higher dimensions.  It was suggested \cite{Kats:2006xp} that the modification occurs in such a way that $(Q/\mu)_{\mathrm{ext}}\geq1$. Finding the nature of HD-corrected solutions in 2D, besides their relevance for a possible formulation of WGC,  is a worthwhile goal in itself,  a result that seems to be missing from the literature.

\subsection{Failure of Perturbation Theory}
We employ perturbation theory to calculate the corrected $(Q/\mu)_{\mathrm{ext}}$ due to the duality-invariant HD terms.  In the function $F(\eig)$ in \eqref{feigs}, we truncate the series at the four-derivative term with $c_4 = \epsilon c$  ($\epsilon\ll1$) and work to $\mathcal{O}(\epsilon)$.  The background fields \eqref{2dersol} receive $\mathcal{O}(\epsilon)$ corrections,
\baa
m = \overline{m} + \epsilon \, \delta m, \quad n = \overline{n} + \epsilon \, \delta n,  \quad V = \overline{V} + \epsilon \, \delta V.\label{pertbg}
\ea
We choose a gauge where the string dilaton is fixed to $\phi=\overline{\phi}$. The dilaton actually \emph{defines} the spatial coordinate $x$: leaving it fixed is analogous to considering the size of the transverse space fixed in higher dimensions.  We can solve for the equations of motion arising from \eqref{actfull} and imposing suitable boundary conditions at asymptotic infinity, $x\to\infty$,  to ensure that the corrections are sub-leading.  The solution is given in eq. \eqref{pertcomps}. 

Rather surprisingly,  perturbation theory breaks down near the horizon. Near the unperturbed event horizon $x=x_+$,  the curvature and electric field strength behave as,  see eq. \eqref{pertfullRF},
\baa
R \sim \frac{\epsilon c\ldc^2}{\qty(x-x_+)^2},  \qquad F^2 \sim \frac{\epsilon c \ldc^3}{x -  x_+},
\ea
which indicates a failure of the perturbative approach near the \emph{non-extremal} horizon.
Such a dramatic breakdown of perturbation theory is disconcerting and surprising from an EFT point-of-view.  Thus,  any hope of systematically studying the effects of HD terms on BHs would appear to be lost.

\subsection{Non-Perturbative Parametrization}
There does exist a framework to obtain a solution in presence of a full tower of HD terms.
In \cite{Moitra:2025fhx},  a non-perturbative parametrization for charged BHs was derived, building on the works \cite{Gasperini:2023tus,Codina:2023nwz}. The solution proceeds from the assumption that the derivative $f(\eig)=F'(\eig)$ of the function appearing in \eqref{feigs} has a single-valued inverse $\eig=\eig(f)$.  In this $f$-parametrization,  the duality-invariant dilaton is simply $\Phi=\log(f/\ldc)+\Phi_0$ and in the gauge $n(x)=1$,  the BH exterior is described by \cite{Moitra:2025fhx}
\baa
x &= x_0 + \int_f^{\infty} \dd{f'} \frac{1}{f' \sqrt{\ldc^2 - G(f') }}, \\
m(f)^2 &= \frac{1}{\mathcal{Q}^2} \csch^2 \qty[ \int_{f_m}^f \dd{f'} \frac{\eig (f')}{\sqrt{2} f' \sqrt{\ldc^2 - G(f') }}  ], \\
V(f) &= \mathcal{Q} \int_0^f \dd{f'} \frac{\eig(f') m(f')^2}{f' \sqrt{\ldc^2 - G(f')}}, \label{fpargensol}
\ea
where $x_0$, $\mathcal{Q}$ and $f_m<0$ are constants and $G(f)=\int_0^f\dd{f'}\eig(f')$. 

For the two-derivative theory, $\eig(f)=-f$.  For an HD extension,  one considers \cite{Codina:2023nwz} a perturbation of the form,
\baa
\eig(f)=-f \pqty{1+h(f^2)},  \label{eigfh}
\ea
where $h(z)$ is a function satisfying $h(0)=0$ and decaying sufficiently fast at infinity.  The function is bounded in a way so that $\ldc^2 - G(f) >0$ between $f=0$ (spatial infinity) and $f=\infty$ (event horizon).  In terms of the function
\baa
\funa(y) \equiv \int_0^y \dd{y'} h(y'),  \label{funadef}
\ea
we have
\baa
G(f) = - \frac{1}{2} \pqty{ f^2 + \funa(f^2)}.  \label{gf2def}
\ea
Thus, $h(f^2) = \funa'(f^2)$.  For a BH solution, $\funa(\infty)\equiv\funa_0$ must be finite \cite{Codina:2023nwz}.  We can now see why the example of the $\eig^4$ term in the previous subsection failed to work.  In this case,  for large $f$, $h(f^2) \sim 1$ and hence $|\funa_0|=\infty$ and there is no BH.  A related example was also studied in \cite{Codina:2023nwz}.

We can now readily make use of the general solution \eqref{fpargensol} to find the extremal charge-to-mass ratio.  It is convenient to express the solution in Schwarzschild-Droste--like gauge \eqref{2dersol}.  To be precise,  we shall use the various transformation formulas to use the string dilaton $2\phi=\Phi+\log m$ as a coordinate, whereby we obtain a line element of the form
\baa
\dd{s}^2 =  - m(\phi)^2 \dd{t}^2 + n(\phi)^2 \frac{4\dd{\phi}^2}{\ldc^2} .  \label{metphi}
\ea

We first perform an expansion about asymptotic infinity. 
The solution is given by \eqref{farmetV}; up to the order shown, it is identical in form to the two-derivative result \eqref{2dersol}.  The HD terms change only the terms sub-leading to \eqref{2dersol}. We thus identify $\mathcal{Q}=\csch\mta$ and the charge-to-mass ratio,
\baa
\frac{Q}{\mu} = \sech\mta, \label{mqcosh}
\ea
where
\baa
\mta \equiv  \int_{f_m}^0 \dd{f'}  \frac{1 + h(f'^2)}{\sqrt{2\ldc^2 + f'^2 + \funa(f'^2) }}. \label{defmta}
\ea

The extremality condition, on the other hand,  must be deduced from a near-horizon ($f=\infty$) expansion,  eq. \eqref{nearhorexp}.  In this case,  assuming $\funa(f^2)$ admits an expansion of the form $\sum_{n=0}^{\infty}\funa_nf^{-2n}$,  (neglecting possible exponentially small  corrections) we find from the near-horizon metric \eqref{metnhexp} the extremality condition to be given by $\Delta = 0$, where
\baa
\Delta \equiv 4\ldc^2 - e^{2(\mta+\mtb)} \qty(\funa_0+2\ldc^2), \label{extrzero}
\ea
with
\baa
\mtb &\equiv \int_0^\ldc \dd{f'} \frac{1 + h(f'^2)}{\sqrt{2\ldc^2 + f'^2 + \funa(f'^2) }} \\
&\quad + \int_\ldc^\infty \dd{f'} \pqty{\frac{1 + h(f'^2)}{\sqrt{2\ldc^2 + f'^2 + \funa(f'^2) }} - \frac{1}{f'}} . \label{defmtb}
\ea

We have to perform a suitable rescaling of the coordinates to transform the metric components \eqref{metnhexp} to a more familiar form.  Note that, in general,  the coordinate system breaks down in the limit $\Delta \to 0$.  This is not unexpected from the discussion in \cite{Moitra:2025fhx}.  (However,  as clarified in the End Matter,  the limit is smooth for exponentially small corrections to $\funa(\infty)$: when $\funa_{n}=0$ for $n>0$.) Since our goal is to find precisely this limiting value,  we need not worry about this possibility.   Setting $\Delta=0$ gives the corrected extremal charge-to-mass ratio,
\baa
\pqty{\frac{Q}{\mu}}_{\mathrm{ext}} = \frac{4\ldc\sqrt{\funa_0 + 2\ldc^2}}{4e^{-\mtb}\ldc^2  + e^{\mtb} (\funa_0 + 2\ldc^2)}.  \label{corrqm}
\ea
This is one of the main results in this Letter,  which expresses the extremal charge-to-mass ratio in terms of the parameters of the theory.  Notice that this relation is state-independent.  By the AM-GM inequality,  we immediately see that irrespective of the form of the corrections,   $\pqty{{Q}/{\mu}}_{\mathrm{ext}} \leq 1$.  This bound is completely general in a duality-invariant theory and cannot be changed by tuning the Wilson coefficients. The two-derivative theory saturates the bound but is not unique in this regard.

We see that the bound is \emph{opposite} of the usual WGC bound in higher dimensions.  This is not too surprising in the present two-dimensional scenario, for reasons mentioned in the Introduction. It would be interesting to further explore related aspects in this set-up.

\section{Attractor Mechanism and Entropy}\label{sec-attrac}

Extremal BHs have also been crucial in our understanding of BH thermodynamics and statistical mechanics,  especially vis-\`{a}-vis state-counting \cite{Strominger:1996sh}.  Extremal BHs often exhibit the attractor mechanism \cite{Ferrara:1995ih},  in which the near-horizon properties of the BH are determined only by the charges carried by it and not by asymptotic moduli.  In the near-horizon region,  the BHs have a two-dimensional Anti-de Sitter ($\mathrm{AdS}_2$) geometry and a covariantly constant electric field. The entropy function formalism \cite{Sen:2005wa, *Sen:2007qy} is an elegant method to calculate the Wald entropy \cite{Wald:1993nt} due to the HD terms in the action.  We now turn our attention to this problem.  The two-dimensional heterotic BH entropy was also considered in \cite{Hyun:2007ii,  *Sadeghi:2007kn} in a specific, limited context with the HD corrections involving only the Ricci scalar.

In this formalism,  one starts with a two-dimensional Lagrangian density (possibly after integrating over the transverse directions,  which does not apply here).  Inserting in this Lagrangian the following \emph{Ans\"atze} for the metric, gauge field and scalar,
\baa
\dd{s}^2 = \fv \pqty{ - x^2 \dd{t}^2 + \frac{\dd{x}^2}{x^2} },  \,\,\, V'(x) &= \fe,\,\, \phi(x) = \phi_0, \label{attran}
\ea
one obtains the function $\mathfrak{f}(\phi_0,  \fv,  \fe)$.  One then defines the Legendre transform \baa
\mathcal{E}(\phi_0,\fv,\fe,\fq)=2\pi\qty(\fe\fq-\mathfrak{f}(\phi_0,  \fv,  \fe)) \label{legenatt}
\ea 
and sets each of $\pdv*{\mathcal{E}}{\fv}$,  $\pdv*{\mathcal{E}}{\phi_0}$ and $\pdv*{\mathcal{E}}{\fe}$ to zero (known as the \emph{attractor equations}).  Inserting the solution in \eqref{legenatt} one obtains the entropy.

It is worth explaining one technical detail at length. The Lagrangian density we use must be supplemented with an additional total-derivative term in comparison with \eqref{actfull},
\baa
\mathcal{L} = n e^{-\Phi} \pqty{\ldc^2 + \frac{\Phi'^2}{n^2} + F\qty(\eig)} -2 \dv{x}\pqty{\frac{e^{-2\phi} m'}{n} }. \label{fulagra}
\ea
While the inclusion of this term does not affect the equations of motion,  it does matter in the entropy computation.  At the two-derivative level,  the generally covariant Lagrangian \eqref{2derac} reduces precisely to the form \eqref{fulagra}.  In obtaining the duality-invariant form \eqref{actfull},  we discard the total derivative \cite{Moitra:2025fhx}.  In the derivation of the HD action \eqref{actfull},  one performs repeated integration by parts \cite{Hohm:2019jgu, *Codina:2023fhy,  Moitra:2025fhx}.  These two procedures are,  however, not on equal footing.  In the second case,  we use the two-derivative equations of motion to perform field redefinitions.  These are consistent with covariant field redefinitions; see,  for example,  \cite{Meissner:1996sa}.  It is known that the entropy evaluated in this formalism is invariant under such field redefinitions \cite{Sen:2005wa, *Sen:2007qy}.

Inserting the \emph{Ans\"atze} \eqref{attran} in \eqref{fulagra},  the suitable combinations of the first two attractor equations yield
\baa
\frac{\fe^2 -2\fv}{2\fv^2} + \sum_{n=2}^\infty c_{2n} \pqty{\frac{2\fv - \fe^2}{\fv^2}}^n - \frac{1 - \fv \ldc^2}{\fv} &= 0,\\
(\fe^2 - \fv)\bqty{1 -2\sum_{n=2}^\infty n c_{2n} \pqty{\frac{2\fv - \fe^2}{\fv^2}}^{n-1}  } - \fv &=0. \label{attreqns}
\ea
It is not hard to see that the equations admit the exact, physically unique solution $\fe^2=2/\ldc^2=2\fv$,. From the third attractor equation,  we also find $e^{-2\phi_0} = |\fq|/(\sqrt{2}\ldc)$,  whereby we find the entropy,
\baa
S = 2\sqrt{2}\pi \ldc^{-1} |\fq|, \label{finent}
\ea
which exhibits no dependence on the Wilson coefficients.  The HD terms therefore do not renormalize the leading order two-derivative entropy.   This non-renormalization is another key result of this Letter.

It is worth pointing out some  interesting features of the entropy.  Note that the $\mathrm{AdS}_2$ radius is of order the string length,  $L_{\mathrm{AdS}}=\sqrt{\alpha'/8}$.  This necessitates that we take into account all HD terms in the effective action.  We implicitly assumed in \eqref{feigs} that the coefficients $c_{2n}$ decay sufficiently fast so that the infinite sum converges at least within some finite radius;  see \cite{Codina:2023nwz}.  Note that the duality-invariant combination $\eig^2$ ensures that non-renormalization persists even if we truncate the sum in \eqref{feigs} to any finite number of terms.  The constant dilaton scenario here is very different from the  linear dilaton case before in that it is possible to find a consistent BH solution even with a finite number of HD terms.  Furthermore, unlike in many typical examples,  the near-horizon electric field $\fe$  is not a free parameter but fixed by the attractor equations.  It would be interesting to examine if the quantization of charge plays any role in this non-renormalization.

We emphasize that the non-renormalization is noteworthy for the reason that we worked in an EFT framework with no input apart from duality invariance.   One is compelled to wonder whether this EFT ``knows'' physics beyond this input.  A relation between the attractor mechanism and topological strings has been conjectured in \cite{Ooguri:2004zv}.  It would be interesting to examine if the entropy is protected because it  describes some topological sector of the theory.

\section{Discussion}\label{sec-discuss}

The different results in the previous sections are intriguing from multiple perspectives.  Upon the introduction of a single duality-invariant HD term,  perturbation theory breaks down rather dramatically, against EFT expectations, and a singularity seems to develop in the vicinity of the BH horizon, originally present in the two-derivative theory.  It would be interesting to examine the formation and dynamical behavior of such a singularity \cite{Moitra:2022umq}.

The situation is ameliorated by the introduction of an infinite number of HD terms,  satisfying certain conditions outlined previously.   Using the non-perturbative form of the solutions,  we calculated an explicit form of the HD correction to the extremal charge-to-mass ratio.  Remarkably, we found a specific bound for this ratio,  albeit one that goes in the opposite direction compared to higher dimensions \cite{Kats:2006xp}. The existence of the bound itself is indicative of universality of some kind: it would be interesting to find similar properties of the two-dimensional string landscape in the spirit of the Swampland program.

In relation to the extremal charge-to-mass ratio, the non-renormalization of the attractor entropy might seem a little puzzling at first. It was argued in \cite{Cheung:2018cwt} that the difference in entropy due to HD terms is non-negative --- a result that must hold in general --- and also proportional to the correction to the charge-to-mass ratio.  While we have $\Delta S=0$,  the second part of the argument does not extend to our case. There is no contradiction because the higher-dimensional derivation relied on perturbation theory,  which we have found to break down in 2D. 

The results can be easily generalized to $r$ Maxwell fields \cite{Moitra:2025fhx}. For instance,  the form of \eqref{corrqm} would be the same with the replacement $Q\to\qty|\vec{Q}|$.  In the entropy calculation,  one finds similarly $\vec{\fe}^2=2/\ldc^2=2\fv$. This further suggests a role of charge quantization in the non-renormalization of the entropy since only one out of $r$ Maxwell fields seems to be relevant to the story. One would have in this case,  similar to \eqref{finent}, $S= 2\sqrt{2}\pi \ldc^{-1} |\vec{\fq}|$.

The non-renormalization of the entropy invites a more refined understanding of the underlying mechanism.  It is known that \cite{Achucarro:1993fd,  *Lowe:1994gt, *Giveon:2004zz} \footnote{I thank Y. Zigdon for pointing out a reference in \cite{Achucarro:1993fd,  *Lowe:1994gt, *Giveon:2004zz}.} that the two-dimensional extremal BH  can be obtained as an orbifold of an extremal $\mathrm{AdS}_3$ BH.  Non-renormalization of the \emph{microscopic} entropy in  $\mathrm{AdS}_3$  has been argued for in \cite{Dabholkar:2006tb}. It would be interesting to see how our low-energy EFT captures such microscopic details irrespective of the values of the Wilson coefficients. It is worth emphasizing that without insisting on duality invariance,  such a non-renormalization would not hold true: generic HD corrections would modify \eqref{finent}.  We have only considered the string tree-level action \footnote{Since the HD action is constructed perturbatively,  one can entertain the possibility that the non-renormalization is non-perturbatively violated.}.  A consideration of the quantum entropy \cite{Sen:2008vm} would shed more light on the microscopic aspects ---  the duality group would also be broken down to a discrete subgroup. In this respect, it might be worth considering the torus partition function of strings on orbifolds of $\mathrm{AdS}_3$ \cite{Dabholkar:2023tzd, *Dabholkar:2025hri}.

Considering a small departure from extremality at the attractor point would be an interesting direction to explore.   While the Jackiw-Teitelboim (JT) model \cite{Teitelboim:1983ux, *Jackiw:1984je}, suitable for describing BHs near extremality \cite{Moitra:2019bub, *Moitra:2018jqs,  *MoitraThesis},  is typically obtained via dimensional reduction,  we can directly obtain the JT model and the corresponding Schwarzian from the two-derivative action in the near-horizon region without any reduction.  
One can combine the results of \cite{Moitra:2019bub, *Moitra:2018jqs,  *MoitraThesis} and \cite{Stanford:2017thb},  and directly obtain the $T^{3/2}$ correction to the BH partition function.  The relative simplicity of two-dimensional string theory would perhaps facilitate comparison between the effective and microscopic theories.  We leave such promising questions for future investigations.

\subsection*{Acknowledgments}

My research is supported by the European Union’s Horizon 2020 Research and Innovation Programme (Grant Agreement No. 101115511). 

\bibliography{hetent_refs}
\onecolumngrid
\phantomsection\label{sec-supp}

\section{End Matter}

\subsection{Perturbative and Non-Perturbative Solutions to the Equations of Motion}

For the perturbative analysis,  as described in the main text,  we consider the HD action   of the form \eqref{actfull} with a single four-derivative term,
\baa
I_{(4)} = \int \dd{x} n(x) e^{-\Phi(x)} \bqty{\ldc^2 + \frac{\Phi'(x)^2}{n(x)^2} - \frac{2m'(x)^2 -V'(x)^2}{2n(x)^2 m(x)^2} + \epsilon c \pqty{\frac{2m'(x)^2 -V'(x)^2}{n(x)^2 m(x)^2}}^2 }, \label{4der-again}
\ea
with $\epsilon\ll1$ a perturbative parameter and the duality-invariant dilaton $\Phi$ related to the string dilaton $\phi$ by \eqref{dinvdil}.  With the $\mathcal{O}\qty(\epsilon^0)$ solution given by \eqref{2dersol}, we consider perturbative \emph{Ans\"{a}tze} of the form \eqref{pertbg}. Solving the resulting equations of motion and suitably choosing constants so as not to change the asymptotically measured mass and charge,  we find
 \baa
\delta m(x) &= \frac{c\ldc^2 e^{-2\ldc x} \qty(e^{\ldc x} \mu - Q^2 )}{\overline{m}(x)} \bqty{ \frac{(2Q^2 - 3\mu^2) e^{-2\ldc x} + 2 e^{-\ldc x} \mu - 1 }{\overline{m}(x)^2} + \frac{e^{\ldc x} -\mu}{\sqrt{\mu^2-Q^2}} \coth^{-1}\frac{e^{\ldc x} -\mu}{\sqrt{\mu^2-Q^2}} }, \\
 \delta n(x) &= \frac{2c\ldc^2 e^{-\ldc x}  \pqty{\mu^2 -Q^2}}{\overline{m}(x)^3} \bqty{ \frac{9e^{-2\ldc x}\qty(e^{\ldc x} -\mu )}{2\overline{m}(x)^2} - 4 e^{-\ldc x } - \frac{1}{2\sqrt{\mu^2-Q^2}} \coth^{-1}\frac{e^{\ldc x} -\mu}{\sqrt{\mu^2-Q^2}}}, \\
\delta V(x) &=  \sqrt{2}c\ldc^2 Q e^{-\ldc x} \bqty{ \frac{(2Q^2 - 3\mu^2) e^{-2\ldc x} + 2 e^{-\ldc x} \mu - 1 }{\overline{m}(x)^2} + \frac{e^{\ldc x} -\mu}{\sqrt{\mu^2-Q^2}} \coth^{-1}\frac{e^{\ldc x} -\mu}{\sqrt{\mu^2-Q^2}} }.
\label{pertcomps}
 \ea
 
 The scalar invariants up to $\mathcal{O}(\epsilon)$ are given by
 \baa
R &=  2 e^{-2\ldc x} \ldc^2 \pqty{e^{\ldc x} \mu - 2Q^2}  - 2 \epsilon c\ldc^4 e^{-\ldc x} \sqrt{\mu^2 - Q^2} \coth^{-1}\frac{e^{\ldc x} -\mu}{\sqrt{\mu^2-Q^2}}  \\
&\quad + 2 \epsilon c\ldc^4 e^{-2\ldc x}\qty(\mu^2 - Q^2) \frac{1 + 21 e^{-\ldc x} \mu- e^{-2\ldc x} (41Q^2 + 36\mu^2) + 87 e^{-3\ldc x}  \mu Q^2 - 32 e^{-4\ldc x} Q^4 }{\overline{m}(x)^4} , \\
F_{\mu \nu}F^{\mu \nu} &= -4 \ldc^2 Q^2 e^{-2\ldc x} - \epsilon c \ldc^4  e^{-4\ldc x} \frac{64  Q^2 \pqty{ \mu^2 - Q^2}}{\overline{m}(x)^2}.  \label{pertfullRF}
 \ea
 Since the red-shift factor $\overline{m}(x)^2$,  with the unperturbed horizon,  is present in the denominator in the scalar invariants, we conclude that perturbation theory breaks down in the vicinity of the horizon.
 
 We next turn to the non-perturbative solution parametrized as \eqref{fpargensol}. In order to determine the mass and charge at asymptotic infinity,  we expand various quantities about $f = 0$, which yields
 \baa
\dv{x}{f} &= - \frac{1}{\ldc f} + \frac{f}{4\ldc^3} - \frac{3 - 4\ldc^2 \funa''(0)}{32\ldc^5} f^3 + \cdots,\\
m(f)^2 &= \frac{\csch^2 \mta}{ \mathcal{Q}^2} \bqty{ 1 - \frac{\sqrt{2} f }{\ldc} \coth \mta + \frac{2+ \cosh 2\mta}{2} \csch^2 \mta \frac{f^2}{\ldc^2} +\cdots} ,\\
2\phi &= \log\frac{f}{\ldc} + \underbrace{\Phi_0 + \log \frac{ \csch \mta}{\mathcal{Q}}}_{\equiv2\phi_1} - \frac{f \coth \mta}{\sqrt{2} \ldc} + \frac{f^2 \csch^2\mta}{4\ldc^2}+\cdots,  \label{smallf}
\ea
where $\mta$ is defined in eq. \eqref{defmta}.

Inverting the last expansion,  we obtain the expansion of the metric and gauge field components in the string dilaton coordinate,
\baa
m(\phi)^2  &= \frac{\csch^2 \mta}{\mathcal{Q}^2} \bqty{ 1 - \sqrt{2} \coth\mta\,  e^{2(\phi-\phi_1)}  + \frac{e^{4(\phi-\phi_1)}}{2} \csch^2\mta + \cdots },  \\
n(\phi)^2 &= 1 + \sqrt{2} \coth\mta \, e^{2(\phi-\phi_1)} + \frac{1+2\cosh 2\mta}{2}\csch^2\mta\,  e^{4(\phi-\phi_1)}+\cdots, \\
V(\phi) &= - \frac{\csch^2\mta}{\mathcal{Q}} e^{2(\phi - \phi_1)} + \cdots.  \label{farmetV}
\ea

In a similar vein,  the determination of the extremality condition requires a knowledge of the metric close to the BH horizon ($f=\infty$).  As mentioned in the main text, we assume the ansatz $\funa(f^2) = \sum_{n=0}^\infty \funa_n f^{-2n}$ and expand various quantities about $f=\infty$,
\baa
\dv{x}{f} &= - \frac{\sqrt{2}}{f^2} + \frac{2\ldc^2 + \funa_0}{\sqrt{2} f^4} - \frac{3 \pqty{2\ldc^2 + \funa_0}^2 -4\funa_1}{4\sqrt{2}f^6} + \cdots, \\
m(f)^2 &= \frac{4\ldc^2e^{-2(\mta+\mtb)}}{\mathcal{Q}^2 f^2} + \frac{2 \ldc^2  e^{-4 (\mta +\mtb)}\Delta}{\mathcal{Q}^2 f^4} + \frac{e^{-6(\mta+\mtb)} \ldc^2 \bqty{ \Delta (5\Delta - 8 \ldc^2)  -12 e^{4(\mta + \mtb) } \funa_1}}{4\mathcal{Q}^2 f^6} + \cdots, \\
2\phi &=\underbrace{ \Phi_0 + \log\frac{2e^{- (\mta +\mtb)}}{\mathcal{Q}} }_{\equiv 2\phi_h} +  \frac{\Delta e^{-2(\mta + \mtb)}}{4f^2} - \frac{12 \funa _1 - e^{-4(\mta+ \mtb) } \Delta(3\Delta - 8\ldc^2) }{32 f^4} + \cdots, \\
 \label{nearhorexp}
\ea
where $\Delta$ and $\mtb$ are defined in  \eqref{extrzero} and \eqref{defmtb} respectively. Then,  inverting the last expression and using the definition of the metric \eqref{metphi},  we find the  near-horizon expansion of the metric components to be
\baa
\mathcal{Q}^2 m(\phi)^2 &= \frac{16\ldc^2}{\Delta} \pqty{\phi - \phi_h} +  8\ldc^2 \frac{\Delta(\Delta+ 8\ldc^2) + 12 e^{4(\mta + \mtb)} \funa_1  }{\Delta^3} (\phi - \phi_h)^2 + \cdots, \\
\frac{\ldc^2}{n(\phi)^2} &= \frac{\Delta (\phi - \phi_h) }{2 e^{2(\mta + \mtb)}}  +  \frac{\Delta(\Delta+ 8\ldc^2) - 36 e^{4(\mta + \mtb)} \funa_1  }{4\Delta e^{2(\mta + \mtb)}} (\phi - \phi_h)^2 + \cdots. \label{metnhexp}
\ea
By making a suitable rescaling of the temporal coordinate,  we can bring $m(\phi)^2$ to the more familiar form in which $m(\phi)^2 \sim \Delta(\phi - \phi_h)$. We can thus find the extremality condition to be $\Delta = 0$. Note that as $\Delta\to 0$, the expansions become ill-defined.  However, in case the large-$f$ corrections to $\funa(f^2)$ are exponentially small,  the limiting behavior as $\Delta\to0$ is smooth.

\end{document}